	\documentclass[aps,showpacs,twocolumn,preprintnumbers,amsmath,amssymb,pra]{revtex4-1}
\usepackage{graphicx}% Include figure files
\usepackage{dcolumn}% Align table columns on decimal point
\usepackage{bm}% bold math
\usepackage{color}
\usepackage{array}
\usepackage{array,multirow,makecell}
\usepackage{graphicx}
\usepackage{float}
\usepackage{tabularx}
\usepackage{color}
\usepackage{colortbl}
\usepackage{xcolor}
\newcolumntype{R}[1]{>{\raggedleft\arraybackslash }b{#1}}
\newcolumntype{L}[1]{>{\raggedright\arraybackslash }b{#1}}
\newcolumntype{C}[1]{>{\centering\arraybackslash }b{#1}}

\begin{document}

\title{Optimal frequency conversion in the nonlinear stage of modulation instability}

\author{A. Bendahmane,$^{1}$ A. Mussot,$^1$ A. Kudlinski,$^{1,*}$ P. Szriftgiser,$^1$ M. Conforti,$^1$  S. Wabnitz,$^2$ and S. Trillo$^3$ }

\email{Corresponding author: alexandre.kudlinski@univ-lille1.fr}

\affiliation{
$^1$PhLAM/IRCICA, CNRS-Universit\'e Lille 1, UMR 8523/USR 3380, F-59655 Villeneuve d'Ascq, France\\
$^2$Dipartimento di Ingegneria dell'Informazione, Universit\`a degli Studi di Brescia and INO-CNR, via Branze 38, 25123 Brescia, Italy\\
$^3$Dipartimento di Ingegneria, Universit\`a di Ferrara, Via Saragat 1, 44122 Ferrara, Italy}

\begin{abstract}
We investigate multi-wave mixing associated with the strongly pump depleted regime of induced modulation instability (MI) in optical fibers.
For a complete transfer of pump power into the sideband modes, we theoretically and experimentally demonstrate that it is necessary to use a much lower seeding modulation frequency than the peak MI gain value. Our analysis shows that a record 95 \% of the input pump power is frequency converted into the comb of sidebands, in good quantitative agreement with analytical predictions based on the simplest exact breather solution of the nonlinear Schr\"odinger equation.
\end{abstract}
													
%\pacs{42.65.Tg}

\maketitle

\section{Introduction}

As well known, MI in a nonlinear dispersive medium described by the nonlinear Schr\"odinger equation (NLSE) leads to the exponential growth of perturbations, at the expense of an intense continuous wave (CW) pump background \cite{VR64plasma,BT66,BF67,YF78,Akh85,Akh86,Tai86}. Since the very beginning of nonlinear wave propagation studies (see Ref. \cite{ZO09} for a review), it has been realized that MI (known also under different names, e.g., Benjamin-Feir instability for deep water waves \cite{BF67}) is a universal pattern generation mechanism in different physical contexts such as plasma physics \cite{VR64plasma}, fluidodynamics \cite{BF67}, and optics \cite{BT66,Tai86}.  Despite fifty years of studies, the investigation of MI is still an extremely active field of interdisciplinary research.

Initial MI studies essentially aimed at the observation of the early stage of exponential growth of the wave perturbation spectrum. It is only recently that the main focus of the experiments has been moved to the fully nonlinear (or long-term) evolution of MI \cite{FPUrec,FPUrec2,FPUrec_spat,DudleyOE09,HammaniOL11a,HammaniOL11b, BendahmaneOL11, ErkintaloPRL11,FPUreclille}. In spite of the valuable pioneering approaches  \cite{YF78,Akh85,Akh86}, such problem remains open even theoretically, at least in its most general formulation, thus continuing to require considerable efforts \cite{ZG13,ZG14,Biondini15}. Indeed, knowing the long-term evolution of MI is a key issue for the understanding of many complex phenomena in physics, such as Fermi-Pasta-Ulam (FPU) recurrence \cite{FPUrec,FPUrec2,FPUrec_spat,HammaniOL11b, FPUreclille}, higher order pulse-splitting \cite{Calini02,Wabnitz10,ErkintaloPRL11}, homoclinic structures \cite{Akh86,Ablowitz90,TW91}, rogue wave formation \cite{Osborne01,rogue1,rogue2}, the development of turbulence \cite{Ablowitz00}, and supercontinuum generation \cite{DGC06}. Since optical pulse propagation in fibers is well described by the NLSE, optical fibers provide a formidable test bed for the experimental investigation of all of these processes.

Whenever MI is induced in the anomalous dispersion regime of a fiber by means of a relatively weak time-periodic perturbation with period $2\pi/\Omega$, the nonlinear evolution of MI leads, via a cascade four-wave mixing (FWM) process, to the appearance of a comb of harmonics at frequencies $\omega_0 \pm n \Omega$, $n=1,2,3, \ldots$ ($\omega_0$ indicates pump frequency). Comb components exhibit a monotonic growth at the expense of the pump, but only up to a characteristic distance, say, $Z_d$: at this point, the pump is maximally depleted. Further on, the power flow is reversed from the sidebands back into the pump, and subsequent cycles of periodic power exchange among the comb waves (or FPU recurrence) are typically observed.

For a given fiber, the actual degree of maximum pump power depletion is strongly affected by the value of the initial modulation frequency $\Omega$, as well as by the pump power $P$. The initial rate of frequency conversion (or MI gain) is well known to peak at a certain frequency, say $\Omega_{pm}$. At this frequency, nonlinear phase matching occurs, i.e. the pump power induced phase shift cancels the linear dispersive mismatch: $|\beta_2| \Omega_{pm}^2=2\gamma P$ ($\beta_2<0$ and $\gamma$ are the fiber dispersion and its nonlinear coefficient, respectively).

However, setting the initial modulation frequency equal to $\Omega_{pm}$ cannot guarantee that optimal frequency conversion from the pump also occurs in a regime where the pump becomes substantially depleted. Indeed, as the pump is depleted, the modulation frequency that satisfies the nonlinear phase matching condition progressively shifts to lower values. The latter argument suggests that the maximum pump depletion may be increased, by setting the initial modulation frequency to a value which is lower than $\Omega_{pm}$. In such situation, instead of immediately mis-matching the mixing process, pump depletion may actually usefully drive the pump and sidebands towards phase-matching, thus effectively increasing the overall conversion efficiency \cite{Cappellini91}. Despite the general fundamental and applicative interest of wave propagation problems described by the NLSE, to our knowledge, the conditions that lead to the optimum transfer of energy from the pump to the sidebands have not been properly clarified, neither theoretically nor experimentally.

In this paper we show that this question can find a simple answer in terms of the so-called Akhmediev breather (AB) solutions of the NLSE \cite{Akh85,Akh86,DudleyOE09,HammaniOL11a,ErkintaloPLA11,BendahmaneOL11}. Moreover, we present a conclusive experimental evidence of this theoretical finding. As a matter of fact, ABs provide a full analytical description of the frequency comb generation process in fibers. As we will show below, cascaded FWM is responsible for a quantitatively important discrepancy with respect to a simple analytical prediction which may result from the truncated three-wave mixing (TWM) approach \cite{TW91,Cappellini91}. Note that the 3-mode truncation was previously used for the modeling of frequency conversion experiments in fibers , carried out in the strongly pump depleted regime \cite{Marhic01,Kylemark06}.

In order to validate the AB approach, in this work we report a careful characterization of the output FWM efficiency as the modulation frequency $\Omega$ is varied. We also perform cut-back measurements, in order to portray the axial evolution of the pump and sidebands. This allows us to demonstrate that a record pump depletion of 95 \% can be achieved, provided that MI is induced by a modulation frequency which is as much as 30\% lower than the phase-matching prediction for an undepleted pump, in excellent quantitative agreement with the AB solution.

\section{Theory}

Let us consider first the optimal frequency conversion problem from a theoretical point of view. Neglecting fiber loss, the dynamics of the depleted stage of MI in our experiments is well described by the NLSE
\begin{equation}\label{nls}
i\frac{\partial u }{\partial z}-\frac{\beta_2}{2}\frac{\partial^2 u}{\partial t^2}+\gamma |u|^2u= 0,
\end{equation}
where $\beta_2$ is group velocity dispersion and $\gamma$ is the nonlinear fiber coefficient. MI is induced by adding a weak in-phase modulation to the CW input pump in Eq. (\ref{nls})
\begin{equation}\label{iv}
u(z=0,t)=\sqrt{P} \left[ \sqrt{\eta_0} + \sqrt{\eta_s} \exp(i \omega t) \right].
\end{equation}
Here $P$ is the total power, $\eta_0$ and $\eta_s=1-\eta_0$ are the pump and the signal input power fractions, and $\omega$ is the pump-signal frequency detuning. The central question that we want to address is the following: what is the seed modulation frequency $\omega$ that guarantees the strongest (optimal) coupling of the input pump power into the signal-idler harmonic pair?
In order to express this condition in universal form, it is convenient to employ dimensionless units, i.e., we assume in Eq. (\ref{nls}) $\beta_2=-1$, $\gamma=1$, $P=1$. This means that distance $z$, time $t$, and the field $u$ are measured in units of the nonlinear length $Z_{nl}=(\gamma P)^{-1}$, of the characteristic time $T_0=\sqrt{|\beta_2| Z_{nl}}$, and of $\sqrt{P}$, respectively.
In this way, the solution of Eqs. (\ref{nls})-(\ref{iv}) depends on a single parameter only, namely, the dimensionless seed frequency $\omega=\Omega T_0 = \Omega \sqrt{|\beta_2| /(\gamma P)}$.
It is well known that, in these units, the linear stability analysis (LSA) of the CW solution of the NLSE (\ref{nls}) yields MI gain for $0 \le \omega \le 2$, with a peak gain at $\omega=\omega_{pm}$,  $\omega_{pm}= \Omega_{pm} T_0 \equiv \sqrt{2}$ being the normalized phase-matching frequency.

\begin{figure}
	\centering
	\includegraphics[width=0.48 \textwidth]{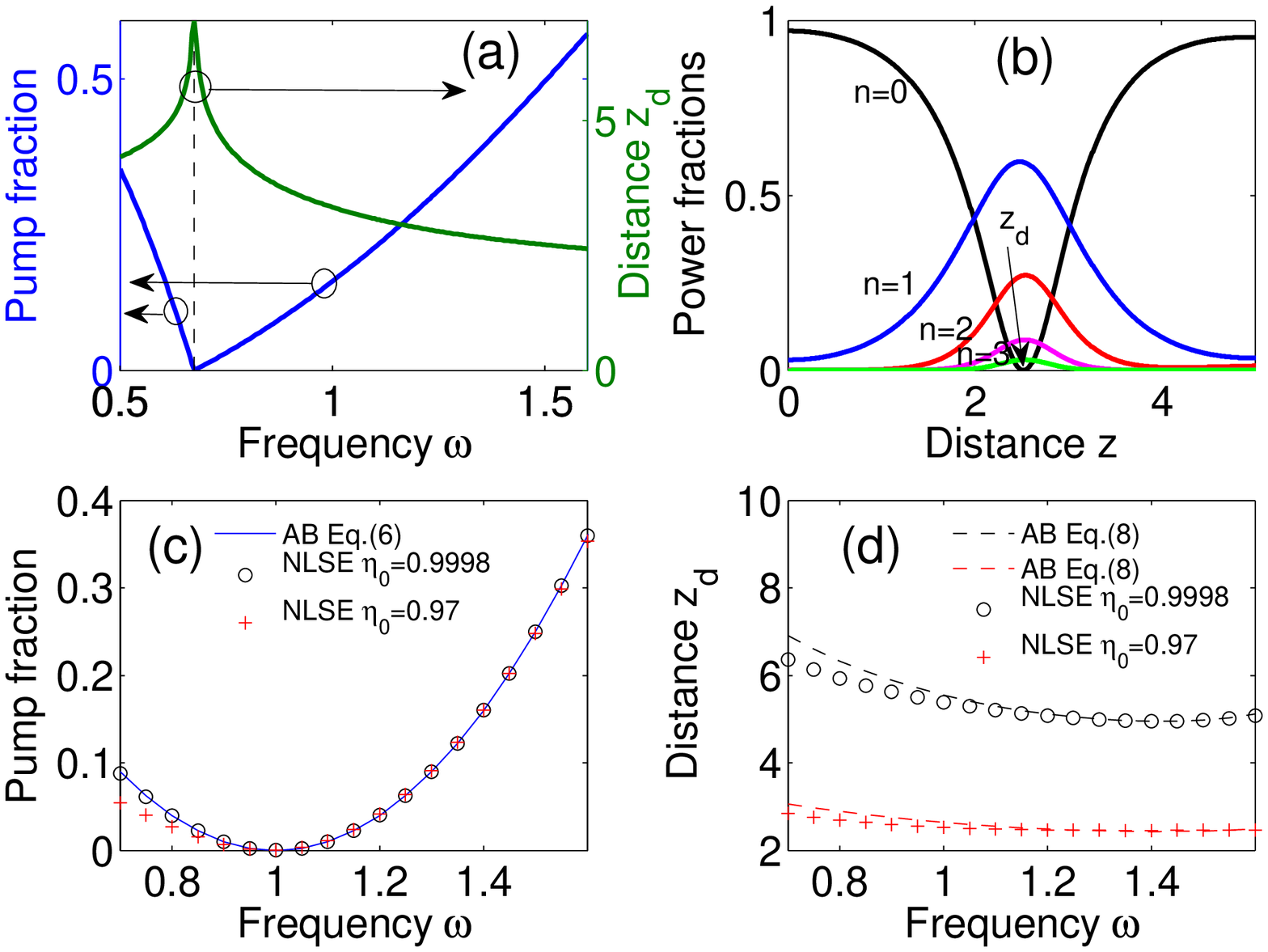}
	\caption{(a) Predictions of the TWM truncated model: Distance $z_d$ and residual pump power fraction $\eta(z_d)$ corresponding to maximally depleted pump versus modulation frequency $\omega$, for a fixed input signal fraction of 3 \% ($\eta_s=0.03$). The dashed vertical line stands for $\omega^{Opt}_{TWM}$ from Eq. (\ref{3wm_fulldepletion}).
	(b,c,d) Full NLSE computation: (b) Evolution of Fourier modal fraction of power at frequency $\omega=\omega^{Opt}_{AB} \equiv 1$;
	(c) Residual pump power fraction vs. $\omega$: AB (Eq. (\ref{pumpAB}), solid line) compared with NLSE simulations (circles and crosses);
	(d) Corresponding distance $z_d$ from NLSE simulations compared with approximation (dashed lines) from Eq. (\ref{distanceAB}).}
	\label{fig1}
\end{figure}

A simple answer to the problem of the optimal condition for frequency conversion from the pump to the sidebands may be obtained by using the truncated (but fully nonlinear) TWM equations for the $n=0, \pm 1$ modes. By substituting the ansatz  $u(z,t)=u_0(z) + u_{1}(z) \exp(-i \omega t) + u_{-1}(z) \exp(i \omega t)$ in Eq. (\ref{nls}), and neglecting higher-order sideband generation, leads to a self-contained set of ordinary differential equations (ODEs) for the pump and the first-order sideband amplitudes $u_0(z), u_{\pm 1}(z)$ \cite{TW91,Cappellini91}. As shown in Ref. \cite{Cappellini91}, such equations can be further reduced, in terms of the variables $\eta=|u_0|^2$ and $\phi=\phi_1+\phi_{-1}-2\phi_0$, to a one-dimensional integrable nonlinear oscillator. The equivalent particle motion is described in the phase plane by the ODEs $d\eta/dz=\partial H/\partial \phi$,  $d\phi/dz=-\partial H/\partial \eta$, with the Hamiltonian
\begin{equation} \label{H}
H = 2\eta \sqrt{(1-\eta)^2-\alpha^2} \cos \phi + (\omega^2+1) \eta - 3 \eta^2/2.
\end{equation}
Here $\alpha=|u_{-1}|^2 - |u_{1}|^2$ is a Manley-Rowe conserved quantity, which reduces to $\alpha=\eta_s$ for the initial condition Eq. (\ref{iv}).
This model predicts that full conversion from the pump to the sidebands occurs whenever the phase plane points corresponding to the the initial condition and to a vanishing pump fraction $\eta=0$ belong to the same level curve of the Hamiltonian, or $H(\eta_0=1-\eta_s)=H(\eta=0)$ in Eq. (\ref{H}). This condition allows us to find a remarkably simple expression for the optimum input modulation frequency (for any given value of the initial power fraction of the signal $\eta_s$)
\begin{equation}\label{3wm_fulldepletion}
\omega^{Opt}_{TWM} = \sqrt{\frac{1}{2} - \frac{3}{2} \eta_s}.
%\omega_{TWM}^2= \frac{1}{2} - \frac{3}{2} \eta_s
\end{equation}
Equation (\ref{3wm_fulldepletion}) predicts that, in the limit case of a small input signal ($\eta_s \ll 1$), the modulation frequency for maximum pump depletion is a factor two lower than the value obtained from the LSA of the undepleted pump, $\omega_{pm}=\sqrt{2}$. Moreover, under the 3-mode approximation, complete conversion is only asymptotically reached (that is, after an indefinitely long propagation distance), since the point $\eta=0$ is a saddle point of the Hamiltonian \cite{Cappellini91}. This situation is summarized in Fig. \ref{fig1}(a): here we display the minimal pump fraction that is obtained at the normalized distance $z_d=Z_d/Z_{nl}$ [i.e., $\eta_{min} \equiv \eta(z_d)$] when using the truncated TWM model. The input power fraction of the signal is equal to 3 \% ($\eta_s=0.03, \eta_0=0.97$). As it can be seen, full pump depletion occurs at the frequency $\omega_{TWM}^{Opt}$ predicted by Eq. (\ref{3wm_fulldepletion}): the corresponding maximum depletion distance $z_d$ diverges to infinity as $\omega_{TWM}^{Opt}$ is approached.

%----------- FIG 1

Equation (\ref{3wm_fulldepletion}) is qualitatively correct in predicting that, for obtaining maximum pump depletion, it is necessary to use a modulation $\omega$ which is substantially lower than the peak LSA gain value. However, this prediction is not quite in quantitative agreement with the NLSE solutions, owing to the presence of higher-order sideband pairs ($n \ge 2$). The coupling of power to the higher harmonics of the initial modulation becomes indeed stronger at modulation frequencies lower than phase-matching frequency, thus affecting also the flow of power towards the primary modes ($n=\pm 1$).
%, i.e., for $\omega \le 1$ NOT INTERESTED IN THSI REGIME
%The presence of sidebands produced by cascade FWM may even reverse, in the case of strong pump depletion, the flow of power towards the primary sideband pair ($n=\pm 1$).

The exact frequency comb dynamics is generally described by doubly periodic (in $t$ and $z$) solutions of the NLSE. However it has been shown that, for $1\le \omega \le 2$, these general solutions remain sufficiently close to the AB solution \cite{Akh85,Akh86} (see also Refs. \cite{Ablowitz90,Osborne01} for a derivation following a different method). The latter is the simplest full solution of the NLSE which is homoclinic to the background, i.e., its associated phase plane trajectory connects the background to itself after a full cycle of evolution (strictly speaking, the AB is a heteroclinic solution, since the trajectory connects two phase shifted background solutions). As it occurs in the TWM model, this cycle of evolution has an infinite period in the longitudinal coordinate $z$. At the apex of the cycle, the AB describes a fully developed train of pulses, which corresponds to a maximally depleted background.

Although the spectrum of the AB solution is symmetric around the pump, recent studies  \cite{DudleyOE09,HammaniOL11a,ErkintaloPLA11,BendahmaneOL11} have shown that ABs may be used to approximate with reasonably good accuracy the nonlinear stage of induced MI with an asymmetric modulation seed, i.e., when a single sideband input condition replaces Eq. (\ref{iv}). This typically requires a small input signal fraction $\eta_s\ll 1$, and relatively large modulation frequencies  (i.e., $1\le \omega \le 2$, so that no harmonics fall under the MI gain bandwidth).

Quite interestingly, we could derive from the AB solutions a simple analytical condition for the optimum input modulation frequency, that leads to maximum pump depletion. Let us expand the AB at its apex (corresponding to the distance $z_d$) in a Fourier series
\begin{eqnarray}\label{AB}
%with a notation
%u_{peak}=\frac{(1-4a) + \sqrt{2a} \cos(\omega t)}{\sqrt{2a} \cos(\omega t) - 1} = \sum_n u_{n,peak} e^{i n 2\pi/\omega},
% omega notation
u_{AB}^{peak}(t) &=& \frac{(\omega^2/2-1) + \sqrt{1- \omega^2/4} \cos(\omega t)}{\sqrt{1- \omega^2/4} \cos(\omega t) - 1} =\nonumber \\
&=& \sum_n \tilde{u}_{n} e^{i n \frac{2\pi}{\omega}t}.
\end{eqnarray}
Here $\tilde{u}_{n}$ are the Fourier coefficients, which can be explicitly calculated. We obtain

%$|\tilde{u}_n|^2=\omega^2 [(2-\omega)/(2+\omega)]^n$.
\begin{eqnarray} \label{pumpAB}
|\tilde{u}^{peak}_{0}|^2=(\omega-1)^2,
\end{eqnarray}
for the pump ($n=0$), whereas for sideband modes $\pm n$,
\begin{eqnarray}\label{sideAB}
|\tilde{u}^{peak}_n|^2=\omega^2 \left( \frac{2-\omega}{2+\omega} \right)^n.
\end{eqnarray}
Equation (\ref{pumpAB}) implies that the pump is totally depleted at $\omega^{Opt}_{AB}=1$. This is confirmed by the numerical simulation of the NLSE (\ref{nls}) with initial condition (\ref{iv}): see Fig. \ref{fig1}(b), where we report the evolution of the power fraction of the pump and the first four sideband pairs. Solving the NLSE at different modulation frequencies $\omega$ confirms that the parabolic law  of Eq. (\ref{pumpAB}) indeed provides a quantitatively accurate description of the maximally depleted pump in the whole range $1 \le \omega \le 2$, regardless of the initial power fraction of the signal. This agreement is displayed in Fig. \ref{fig1}(c), where we compare the results of NLSE simulations (with two different input signal fractions $\eta_s=2 \times 10^{-4}, 300\times 10^{-4}$, or pump fractions $\eta_0=0.9998$ and $\eta_0=0.97$),
%$\eta_0=1-2\times 10^{-4}$ and $\eta_0=1-300\times 10^{-4}$ or $\eta_s=2 \times 10^{-4}, 300\times 10^{-4}$)
to the analytical expression [Eq. (\ref{pumpAB})]. As it can be seen, slight discrepancies only appear for significantly high input signal fractions (see crosses for $\eta_s=0.03$), and in the range of modulation frequencies well below the optimum value $\omega^{Opt}_{AB}=1$.

Note also that, although the input signal fraction does not significantly affect the amount of maximum pump depletion (at least for $1 \le \omega \le 2$), $\eta_s$ strongly affects the maximum pump depletion distance $z_d$. The AB solutions also provide a reasonably good estimate for $z_d$ in terms of the following formula \cite{ErkintaloPLA11} (an alternative formula is reported in Ref. \cite{Osborne01})
\begin{equation}\label{distanceAB}
z_d=\frac{1}{\omega \sqrt{1- \omega^2/4}} \ln \left( \sqrt{  \frac{\omega (1-\omega^2/4)}{\eta_s}  } \right).
\end{equation}
Fig. \ref{fig1}(d) shows that Eq. (\ref{distanceAB}) is indeed in good agreement with the numerical solution of Eqs. (\ref{nls}-\ref{iv}).

In summary, based on the AB solutions, one predicts that total pump depletion occurs precisely at $\omega^{Opt}_{AB}=1$. Such modulation frequency is higher than the prediction of the 3-mode truncation (i.e., $\omega_{TWM} \simeq 1/\sqrt{2}$), but it is still substantially lower (by a factor $\sqrt{2}$) than the phase-matching frequency
$\omega_{pm} =\sqrt{2}$.
% yet in the range where no additional harmonics are linearly unstable.

There is another difference between the results of the 3-mode truncation and the AB theory which is worth to emphasize. In the former case, the condition for optimal pump depletion necessarily coincides with the condition for maximum power in the sideband modes. Conversely, the AB solution implies that the optimum power conversion from the pump into, e.g., the primary sidebands occurs at the modulation frequency $\omega^{Opt}_{n=1}=\sqrt{5}-1 \simeq 1.24\neq \omega^{Opt}_{AB}$. This frequency is obtained by maximizing the power fraction $|\tilde{u}_1|^2$ in Eq. (\ref{sideAB}) [in general each sideband order $n$ peaks at a different frequency, which can be easily calculated from Eq. (\ref{sideAB})]. This is because at $\omega^{Opt}_{AB}=1$ higher-order sidebands concur to the full depletion of the pump [even if $\sim 96 \%$ of the power is contained in the modes with $n \le 3$, the fractional content of $n=2,3$ modes is far from being negligible, see also Fig. \ref{fig1}(b)]. Viceversa, by slightly increasing the frequency above $\omega^{Opt}_{AB}=1$, the AB spectrum narrows down (i.e. the sideband modes decay faster for increasing $n$). In this case, even if the pump is not depleted completely, the fraction of power in the $n=1$ sideband pair may still reach its largest value.

%----------- FIG 2
\begin{figure}
	\centering
		\includegraphics[width=0.45 \textwidth]{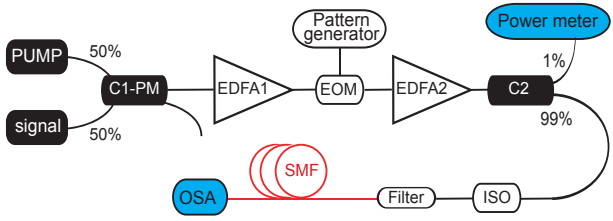}
	\caption{Experimental set-up. C1-PM, polarization maintaining coupler; EDFA, erbium-doped fiber amplifier; EOM, electro-optic modulator; C2, coupler; ISO, isolator; SMF, Corning 1.1 km single mode fiber; OSA, optical spectrum analyzer. }
	%The inset shows the spontaneous spectrum of MI recorded at the fiber output.
		\label{fig2}
\end{figure}

\section{Experimental results}
In order to experimentally investigate the optimal conditions for pump depletion in the MI process, we employed the set-up reported in Fig. \ref{fig2}.
To allow for relatively large modulation frequencies, we induced MI by an asymmetric seed. A strong CW-pump ($\lambda=1560$ nm) and a weak tunable CW-probe were combined and intensity modulated to generate 12 ns square pulses at 5 MHz repetition rate. For increasing the pump pulse peak power, we used an erbium-doped fiber amplifier (EDFA2) before launching the modulated pump into a 1.1 km long Corning SMF28 fiber. For such a relatively short fiber length, the presence of fiber loss can be safely neglected.

In order to fully characterize the MI dynamics, we performed two different sets of experiments. In the first set we measured at the fiber output the residual pump and signal power fractions, as the seed frequency detuning (or modulation frequency) $\Delta f=\Omega/2\pi$ was varied. We compared such results with the measured MI gain curve. In the second set of experiments, we performed cut-back measurements and recorded the full output spectra every 20 m.

%----------- FIG 3
\begin{figure}
	\centering
	\includegraphics[width=0.5 \textwidth]{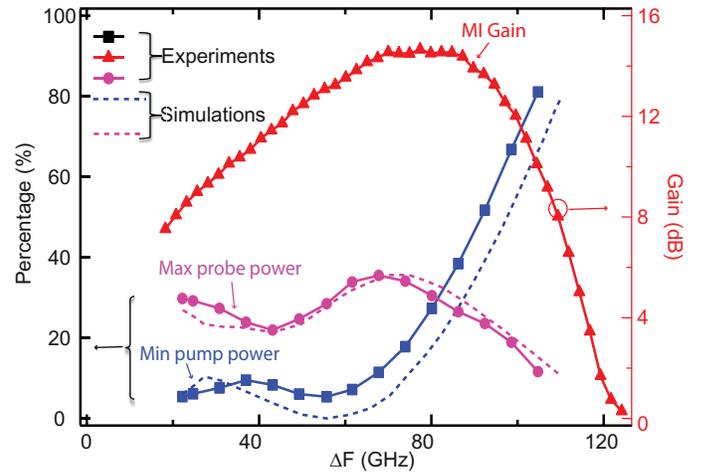}
	\caption{Symbols: experiments. MI gain (red triangles), residual pump (blue squares) and signal (magenta circles) fractions versus frequency detuning $ \Delta f=\Omega/(2\pi)$. Dashed curves: simulations. The same colour code has been used.}
	\label{fig3}
\end{figure}
Fig. \ref{fig3} illustrates the dependence of the fraction of total output power in the pump and in the signal, respectively, vs. their frequency separation. By using an extremely weak signal, the (undepleted pump) MI gain curve (red triangles in Fig. \ref{fig3}) was also measured. The MI gain profile shows that the phase matching frequency is $\Delta f_{pm} \simeq 80$ GHz, in good agreement with the theoretical prediction ($\simeq 84$ GHz). Next we kept the pump power ($P=2.83$ W) unchanged, and we increased the signal power in order to investigate the MI process in the strongly depleted pump regime. In this case, the signal power was accurately adjusted, so that the the fiber length was kept nearly equal to the maximum depletion length $Z_d$. As shown in Fig. \ref{fig3}, the data (blue squares) clearly indicate that a minimum residual pump power of 5 \% (95 \% depletion) was achieved at $\Delta f = 55.7$ GHz. This frequency is in good quantitative agreement with the estimate $\omega^{Opt}_{AB}=1$, which corresponds, in real world units, to $\Delta f = \Delta f_{pm}/\sqrt{2}=59.4$ GHz. The peak of the output signal fraction (magenta circles) was observed at $\Delta f =67.8$ GHz, a value which agrees remarkably well with the predicted value $\omega^{Opt}_{n=1}=1.24$, or $\Delta f = \Delta f_{pm}/1.24=67.7$ GHz. For the sake of clarity, all of our results are summarized in Table 1.
% TABLEAU
\begin{table}[h]
\caption{Results summary}
\label{my-label}
\resizebox{8.5 cm}{!}{%
\begin{tabular}{|c|c|c|c|c|c|}
\hline
\multicolumn{2}{|l|}{\cellcolor[HTML]{9B9B9B}} &  & \multicolumn{2}{c|}{} &  \\
\multicolumn{2}{|l|}{\cellcolor[HTML]{9B9B9B}} &  & \multicolumn{2}{c|}{\multirow{-2}{*}{\begin{tabular}[c]{@{}c@{}} \textbf{Optimal pump} \\ \textbf{conversion frequency}\end{tabular}}} &  \\ \cline{4-5}
\multicolumn{2}{|l|}{\multirow{-3}{*}{\cellcolor[HTML]{9B9B9B}}} & \multirow{-3}{*}{\begin{tabular}[c]{@{}c@{}}\textbf{Perfect}\\  \textbf{phase}\\  \textbf{matching}\end{tabular}} & \textbf{TWM} & \textbf{AB} & \multirow{-3}{*}{\begin{tabular}[c]{@{}c@{}}\textbf{Optimal signal}\\ \textbf{conversion} \\ \textbf{frequency}\end{tabular}} \\ \hline
\multicolumn{2}{|c|}{\textbf{Normalized units}} & $\omega_{PM} =  \sqrt{2}$                                     & $\omega^{Opt}_{TWM}\simeq \dfrac{1}{\sqrt{2}}$                       & $\omega^{Opt}_{AB}=1$                            & $\omega^{Opt}_{n=1}\simeq 1.24$                                                                                                  \\ \hline
 &  &  & \multicolumn{2}{c|}{} &  \\
 & \multirow{-2}{*}{\textbf{\begin{tabular}[c]{@{}c@{}}Measured\\ (GHz)\end{tabular}}} & \multirow{-2}{*}{80} & \multicolumn{2}{c|}{\multirow{-2}{*}{55.7}} & \multirow{-2}{*}{67.8} \\ \cline{2-6}
 &  &  &  &  &  \\
\multirow{-4}{*}{\textbf{\begin{tabular}[c]{@{}c@{}}Physical\\ Units\end{tabular}}} & \multirow{-2}{*}{\textbf{\begin{tabular}[c]{@{}c@{}}Calculated\\ (GHz)\end{tabular}}} & \multirow{-2}{*}{84} & \multirow{-2}{*}{\begin{tabular}[c]{@{}c@{}}$\Delta f_{PM}/2$\\ =~~42\end{tabular}} & \multirow{-2}{*}{\begin{tabular}[c]{@{}c@{}}$\Delta f_{PM}/\sqrt{2}$\\ =~~59.4\end{tabular}} & \multirow{-2}{*}{67.7} \\ \hline
\end{tabular}
}
\end{table}

In order to fully validate our measurement results, we compared them with numerical solutions of the generalized NLSE (GNLSE, which is typically used to simulate broaband pulse propagation and supercontinuum generation \cite{DGC06}), using the available fiber data ($\beta_2= 24.3~ps^2/km$, third-order dispersion $\beta_3= 0.14~ps^3/km, \gamma= 1.2 W^{-1}km^{-1}$, and the loss coefficient $\alpha=0.2~dB/km$) plus a CW pump and seed pair. The outcome of our simulations is superimposed (blue and magenta dashed lines) to the data in Fig. \ref{fig3}. As can be seen, an excellent agreement is obtained for the value of the optimal maximum depletion frequency.

However, at variance with the experimentally observed minimal residual pump of 5\%, simulations predict that full depletion should be observed (see also Fig. \ref{fig1}). Additional numerical simulations matching the pulsed (as opposed to CW) input pump show that the small residual pump fraction is due to the finite roll-off factor of the non ideal square pump pulses. Indeed, averaging over the pump power profile is well known to lead to incomplete frequency conversion in the MI process, as previously reported in, e.g., Ref.~\cite{FPUrec2}.

%----------- FIG 4
\begin{figure}
	\centering
	\includegraphics[width=0.5 \textwidth]{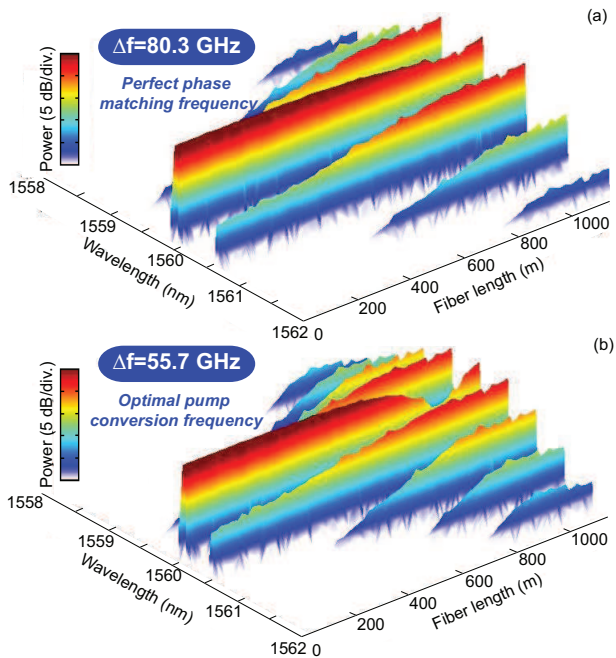}
	\caption{Observed spectra versus wavelength and distance, as reconstructed from cut-back measurements: (a) $f=80.3$ GHz; (b) $f=55.7$ GHz.}
	\label{fig4}
\end{figure}

In our second set of experiments, we recorded the evolution along the fiber length of the pump and signal power by means of a cut back experiment in steps of $20$ m. These results are displayed in Fig. \ref{fig4} (log vertical scale) for two specific values of the pump-signal frequency shift. The first value is close to the perfect phase matching frequency ($\Delta f=80.3$ GHz, Fig. \ref{fig4}(a)), while the second value is close to the optimal conversion frequency ($\Delta f=55.7$ GHz, Fig. \ref{fig4}(b)). As we can see,  higher-order sidebands, that start to grow at longer distances when compared with the primary sidebands ($n=1$), play a non-negligible role around the point of maximal depletion. In addition, Fig. \ref{fig4}(b) shows that pump depletion is more pronounced  around $1000$ m.

\begin{figure}
	\centering
	\includegraphics[width=0.5 \textwidth]{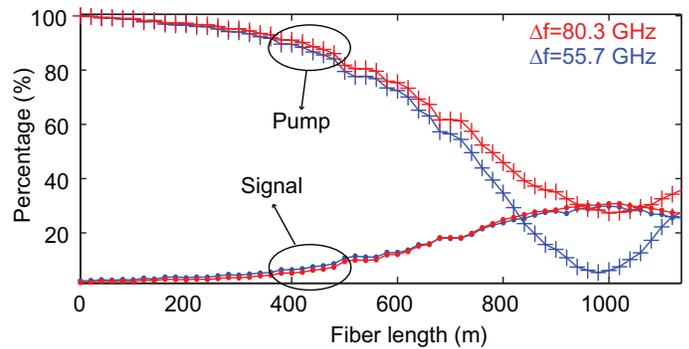}
	\caption{Measured pump and signal fractions against distance obtained from cut-back measurements, for frequencies $f=80.3$ GHz (close to peak linear gain)
	and $f=55.7$ GHz (optimum frequency for depletion), respectively.}
	\label{fig5}
\end{figure}

To gain a clearer insight into the longitudinal variation of the pump and signal power fractions, Fig. \ref{fig5} compares their variation for the two different modulation frequencies of Fig. \ref{fig4}. Fig. \ref{fig5} shows that, for $\Delta f=80.3$ GHz, a relatively large (i.e., $27\%$) residual pump power fraction is observed at $Z_d \simeq 1000$ m. Conversely, data obtained at $\Delta f=55.7$ GHz show that a marked enhancement of the maximal pump power depletion occurs at $Z_d \simeq 980$ m, in good agreement with the theoretical prediction.

%----------- FIG 5rogue

\section{Summary}
In conclusion, we reported a clear experimental evidence that optimal power transfer from a pump wave to its sideband modes, as described by the fully nonlinear stage of the MI process in optical fibers, occurs for an initial modulation frequency $ \Delta f=\Delta f_{pm}/\sqrt{2}$ that is well below the nonlinear phase-matching frequency $\Delta f_{pm}$ where the MI gain peaks. This result is important as it establishes a fundamental property of the nonlinear evolution of modulation instabilities in nonlinear dispersive media. Our observations have also widespread applications to optical frequency conversion devices. Indeed, as it was earlier pointed out for vector MIs in birefringent fibers \cite{cap91}, and confirmed experimentally \cite{trillo97,millot98,seve00}, the concept of the phase-matching of parametric mixing processes is of limited use in the strongly depleted pump regime, unless it is suitably extended by means of a fully nonlinear large-signal theory.

\section*{Funding Information}

This work was partly supported by the ANR FOPAFE (ANR-12-JS09-0005), TOPWAVE (ANR-13-JS04-0004) and NoAWE (ANR-14-ACHN-0014) projects, by the "Fonds Europ\'{e}en de D\'{e}veloppement Economique R\'{e}gional", the Labex CEMPI (ANR-11-LABX-0007) and Equipex FLUX (ANR-11-EQPX-0017) through the "Programme Investissements d'Avenir". Funding from Italian Ministry of Research (grant PRIN 2012BFNWZ2) is also gratefully acknowledged.

%\section*{Acknowledgments}
%Formal funding declarations should not be included in the acknowledgments but in a Funding Information section as shown above. The acknowledgments may contain information that is not related to funding:

%\section*{Supplemental Documents}
%\emph{Optica} authors may include supplemental documents with the primary manuscript. For details, see \href{http://www.opticsinfobase.org/submit/style/supplementary-materials-optica.cfm}{Supplementary Materials in Optica}. To reference the supplementary document, the statement ``See Supplement 1 for supporting content.'' should appear at the bottom of the manuscript (above the references).

%\pagebreak

\end{document}